\documentclass[smallcondensed,final]{svjour3} 
\usepackage{amssymb,amsmath} 
\usepackage{mathptmx} 
\usepackage{graphicx,subfig} 
\usepackage[utf8]{inputenc}

\begin{document} 
\title{Bootstrap percolation and kinetically constrained models on hyperbolic lattices}

\author{ François Sausset \and Cristina Toninelli \and Giulio Biroli \and Gilles Tarjus }

\institute{ F. Sausset \and G. Biroli \at Institut de Physique de Théorique, CEA, CNRS, URA 2306, F-91191 Gif sur Yvette, France \\\email{sausset@lptmc.jussieu.fr} \\\email{giulio.biroli@cea.fr} \and C. Toninelli \at Laboratoire de Probabilités et Modèles Aléatoires, CNRS UMR 7599, Université Pierre et Marie Curie et Université Denis Diderot, 4 Place Jussieu, 75252 Paris Cedex 05, France \\\email{cristina.toninelli@upmc.fr} \and G. Tarjus \at Laboratoire de Physique Théorique de la Matière Condensée, Université Pierre et Marie Curie, UMR CNRS 7600, 4 place Jussieu, 75252 Paris Cedex 05, France \\\email{tarjus@lptmc.jussieu.fr}} 

\date{\today}

\maketitle
\begin{abstract}
	We study bootstrap percolation (BP) on hyperbolic lattices obtained by regular tilings of the hyperbolic plane. Our work is motivated by the connection between the BP transition and the dynamical transition of kinetically constrained models, which are in turn relevant for the study of glass and jamming transitions. We show that for generic tilings there exists a BP transition at a nontrivial critical density, $0<\rho_c<1$. Thus, despite the presence of loops on all length scales in hyperbolic lattices, the behavior is very different from that on Euclidean lattices where the critical density is either zero or one. Furthermore, we show that the transition has a mixed character since it is discontinuous but characterized by a diverging correlation length, similarly to what happens on Bethe lattices and random graphs of constant connectivity. 
\end{abstract}

\section{Introduction} \label{sec:introduction}

The effect of geometry, and more specifically of the curvature of the embedding space, on phase transitions in the context of statistical mechanics or quantum field theory has attracted continued interest among mathematicians and physicists. In the present work, we focus on bootstrap percolation on hyperbolic lattices obtained by regular tilings of the hyperbolic (negatively curved) plane. Bootstrap percolation (BP) with \textit{blocking parameter} $m$ is defined by the following deterministic evolution: starting from a configuration in which each lattice site is independently occupied with a given probability $\rho$, occupied sites are emptied if they have fewer than $m$ occupied neighbors and the process is repeated until reaching a configuration that can no longer be evolved, \textit{i.e.} one which only contains empty sites and occupied sites with $m$ or more occupied neighbors. A site is said \textit{blocked} for the initial configuration if it is occupied in the corresponding final configuration. We investigate the occurrence and the properties of infinite clusters of blocked sites when varying the initial density $\rho$.

This study has both mathematical and physical motivations. Concerning the former, it connects to the series of studies on phase transitions on ``nonamenable graphs''~\cite{Lyons:2000,Schonmann:2002,Balogh:2006}, nonamenability being a notion that informally speaking refers to graphs with a nonvanishing surface (boundary) to volume (bulk) ratio. On the more physical side, it is driven by a line of research on phenomena such as glass and jamming transitions in liquids and related systems. There is an admittedly long route from BP on hyperbolic lattices to glassforming liquids, and the (heuristic) connection that goes through notions such as ``geometric frustration'' in condensed matter and schematic descriptions of slow dynamics in terms of ``kinetically constrained models'' will be discussed in more detail below.

We show that, for a large choice of tilings and blocking parameters, there is a nontrivial critical density $0< \rho_c < 1$ such that infinite clusters of blocked sites only occur when the density $\rho$ of the occupied sites of the initial configuration satisfies $\rho\geq \rho_c$. Although we resort to rather casual physicists' language in the present article, the proof is rigorous and will be presented with due formalism and rigor in a subsequent publication~\cite{toninelli09}. Furthermore we present strong evidence that this transition has a discontinuous order parameter: the probability that a given site is blocked displays a jump at $\rho_c$. Similarly to its counterpart on tree-like structures such as Bethe lattices and random graphs of constant connectivity~\cite{Iwata:2009,Schwarz:2006}, this transition has features of both first-order and second-order phase transitions, namely a discontinuity in the order parameter and a diverging correlation length.

Note that there could be another percolation threshold at a higher density $\rho_u > \rho_c$ such that for $\rho_c<\rho<\rho_u$ there is an infinite number of disconnected (infinite) clusters of blocked sites with a unique cluster occurring only above $\rho_u$. This second transition indeed occurs for site percolation on hyperbolic lattices and nonamenable graphs~\cite{Benjamini:2001,Lalley:1998,Lalley:2001}. It also takes place for BP on Cayley trees and in this case it can be thought of as a boundary-induced phenomenon. We do not investigate this aspect for BP on hyperbolic lattices since it is of less interest in the context of glassforming systems.

The paper is organized as follows. In section~\ref{sec:glass_transition_and_hyperbolic_plane}, we outline the arguments that support the possible relevance of BP on hyperbolic lattices to the study of glass and jamming transitions and we establish the connection between BP and kinetically constrained models for glassy dynamics. In section~\ref{sec:hyperbolic_lattices}, we give basic information concerning the hyperbolic plane and hyperbolic lattices and we recall how one can construct appropriate trees embedded in the hyperbolic lattices. Section~\ref{sec:bootstrap_existence} contains the main analytical result, \textit{i.e.} the existence of a BP transition on hyperbolic lattices, and a sketch of the proof while section~\ref{sec:bootstrap_order} contains our results concerning the order of the transition. The latter are obtained by combining analytical results on Cayley trees, conjectures on the effect of boundary conditions for BP on nonamenable graphs and a numerical study. Finally, we give some concluding remarks, and a more technical point is discussed in the appendix.

\section{Glass transition, kinetically constrained models and hyperbolic plane} \label{sec:glass_transition_and_hyperbolic_plane}

\subsection{Glass formation and geometric frustration} \label{sub:glass_frustration}

As already mentioned, one motivation for the present study comes from phenomena associated with glass and jamming transitions. The glass transition of liquids, colloids and polymers refers to the formation under cooling or compressing of an amorphous phase which can no longer flow on the observation time scale and then appears as a solid. A related phenomenon exists in dense granular assemblies and in foams as these systems get clogged or jammed when the external drive that made them move and flow becomes too small.

Among the numerous theoretical approaches proposed to explain glass formation and jamming, one relates the observed slowing down of dynamics to the presence of ``geometric frustration''~\cite{Sadoc:1999,Nelson:2002,Tarjus:2005}. The latter describes a competition between global and local constraints whose manifestation is an impossibility to tile the whole space by extending the local structure characteristic of the liquid (or of any other glassforming system). Along these lines, a minimal model is provided by a two-dimensional atomic fluid of disks embedded in the hyperbolic plane~\cite{Rubinstein:1983,Sausset:2008b,Modes:2007}. The negative curvature of space prevents the extension of the local hexagonal order and generates frustration. It was found in a computer simulation study~\cite{Sausset:2008b} that as the liquid is cooled, crystallization in hexagonal or related structures is avoided whereas the relaxation time increases rapidly and leads at low temperature to glass formation when the system no longer equilibrates on simulation time scales. At low enough temperature, it was observed that the dynamics is dominated by rare curvature-induced defect structures that are separated on average by a distance of the order of the ``radius of curvature'' $\kappa^{-1}$ of the embedding space (the gaussian curvature of the hyperbolic plane is $K=-\kappa^{2}$).

Studying the dynamics of the glassforming liquid on the hyperbolic plane at lower temperature and following the possibly collective kinetics of the defects is in practice out of reach for computer simulation. To get some insight, one possible strategy is to resort to an even cruder model description: this brings us to the so-called ``kinetically constrained models''. Such models form the core of another line of thought on glassy systems, one that envisages glass formation as a purely kinetic phenomenon with little or no thermodynamic and structural input~\cite{Garrahan:2003}. Slowing down of the relaxation to equilibrium as temperature is decreased is attributed to the emergence of kinetic constraints that restrain the dynamics of the system. Those constraints are introduced in an explicit but \textit{ad hoc} way in a series of lattice models. Before introducing the more specific model we have in mind, it is worth stressing that the collective nature of the dynamics captured in the above simulations and the one which is describable by kinetically constrained models on the hyperbolic lattice involve different scales: roughly below the radius of curvature $\kappa^{-1}$ for the former (due to limited computer resources) and above for the latter (as a consequence of the hyperbolic metric, the basic step of the tiling is always at least of order $\kappa^{-1}$, see Section~\ref{sec:hyperbolic_lattices}).

\subsection{Kinetically constrained models and bootstrap percolation} \label{sub:KCM_and_bootstrap}

The kinetically constrained model we focus on is the one introduced by Fredrickson and Andersen (FA) in~\cite{Fredrickson:1984}. Given a graph $G$ with vertex set $V$ and edge set $E$, one has on each vertex $i\in V$ an Ising spin variable, $\sigma_i=\pm1$.

The Hamiltonian is trivial, $ H = -\frac{1}{2}\sum_i\sigma_i$, \textit{i.e.} the spins do not interact and, hence, the equilibrium distribution is a product measure. On the other hand, the Monte Carlo spin dynamics is subjected to a kinetic constraint: at each time step a randomly chosen spin, say $i$, is flipped with the standard rate $ w(\sigma_i\rightarrow -\sigma_i)={\mbox{min}} \left\{ 1,\, {\rm e}^{-\beta \sigma_i} \right\} $, {\it if and only if} the number of nearest neighbors (\textit{i.e.} vertices connected by an edge to $i$) which are in the state $-1$ is larger than or equal to a fixed value $f$, which is called the {\it facilitation parameter} ($\beta=1/T$ is the inverse temperature, with the Boltzmann constant set to $1$). Physically, plus and minus spins can be thought of as describing small regions of low and high mobility in a glassforming system.

The FA model is one of the most studied kinetically constrained models. On Cayley trees and on Bethe lattices of constant connectivity $q$, the FA model displays a dynamical transition for $1<f<q-1$ \cite{Sellitto:2005,Reiter:1992,Pitts:2000} which is very similar to the ideal discontinuous transition described by the Mode Coupling Theory of glassforming liquids \cite{Gotze:1992,Bouchaud:1996} while for $f=q-1$ a transition takes place with an ergodicity breaking parameter equal to zero. On the other hand, on any finite-dimensional hypercubic (Euclidean) lattice,  this dynamical transition is wiped out by the diffusion of very rare cooperative defects \cite{Reiter:1991,Toninelli:2004,Toninelli:2005}, thus supporting the ``mean-field'' character of the transition. Despite the absence of a true dynamical transition, the FA model on Euclidean lattices displays an interesting glassy behavior including a faster than Arrhenius growth of the relaxation times as the temperature goes to zero when $f>1$~\cite{Ritort:2003}. The mechanism which is responsible for this strong slowing down is that, as the temperature is lowered, $-1$ spins become rarer while the typical size of the regions which should be cooperatively rearranged in order to satisfy the constraint at a given site becomes larger. Here, we analyze the behavior of the FA model on hyperbolic lattices. 

Remarkably, the study of the dynamical glass transition of the FA model can be reduced to the analysis of a static phase transition called bootstrap percolation (BP)~\cite{Chalupa:1979}. In BP one starts from an initial configuration in which each vertex of the graph $G$ is occupied by a particle with probability $\rho$ and empty with probability $1-\rho$. Then, one randomly removes particles that have fewer than $m$ neighboring occupied sites; thus $m$ is called the \textit{blocking parameter}. Iterating this procedure leads to two possible asymptotic results: either the lattice is completely empty, or the remaining occupied sites are all mutually blocked. We let $\rho_c(G,m)$ denote the critical threshold above which blocked clusters occur. We recall that when $G$ is a Euclidean lattice, $\rho_c=1$ for $d < m\leq 2d$, and $\rho_c=0$ otherwise~\cite{Enter:1987,Schonmann:1992}. On the other hand, on infinite Cayley trees, Bethe lattices and random graphs of connectivity $q$, $\rho_c=1$ if $m=q$, $\rho_c=0$ if $m=1$, and $0<\rho_c<1$ if $1<m<q$~\cite{Chalupa:1979}.

The connection with the FA model emerges by noticing that if the graph has constant connectivity $q$, as the lattices we focus on, the BP procedure can equivalently be formulated by saying that one removes a particle if it has at least $f=q+1-m$ empty neighbours\footnote{If the graph does not have constant connectivity and one fixes the facilitation parameter, the procedure does not correspond to one with fixed blocking parameter.}. In addition, one has to identify plus (minus) spins with filled (empty) sites, $\rho$ with the probability for a spin to be equal to $+1$, \textit{i.e.} $\rho=1/(1+\exp(-\beta))$, and $m$ with $m(f) = q-f+1$. Then, it is easy to verify that if for a given initial configuration, the BP procedure ends up with the whole lattice emptied, there exists for the corresponding spin configuration a sequence of spin flips, each satisfying the kinetic constraint, that brings the configuration to that with all spins in the $-1$ state. As a result, all spins can eventually flip if $1/(1+\exp(-\beta))<\rho_c$. Although this is not a guarantee of ergodicity in general, it is so for KCM's, as proved in~\cite{Cancrini:2008}. (The underlying reason is that because of their trivial thermodynamic measure, the only ergodicity breaking that can take place is a ``reducibility transition'' at which the configuration space becomes disconnected in several pieces, each piece containing roughly the same number of configurations: see~\cite{Toninelli:2004a,Toninelli:2008} for a more detailed discussion.) Conversely, if the BP ends up with a nonempty cluster of blocked sites, the latter is by construction formed by all the plus spins that will never move under FA dynamics. Thus, the occurrence of a BP transition is a necessary and sufficient condition for the existence of an ergodicity breaking transition in the corresponding FA model. Furthermore, if the BP transition is discontinuous, \textit{i.e.} if the fraction of blocked sites is strictly positive at $\rho_c$, the dynamical transition for FA is also discontinuous. One can indeed show~\cite{Toninelli:2004a,Toninelli:2008} that, due to the occurrence of a finite fraction of blocked spins, the local time-dependent spin correlation function relaxes to a finite plateau at $\rho_c$ (while below $\rho_c$ it relaxes to zero).

In conclusion, the existence of a dynamical transition for the FA model and some properties of this transition can be studied by focusing on the much simpler static problem of the BP. This is what we have done for hyperbolic lattices. We postpone a direct investigation of the detailed dynamic behavior of kinetically constrained models on such lattices for future work, but it is worth pointing out that the nature of the transition (continuous or first-order) has strong implications for the dynamics in both the ergodic and the nonergodic phases. For instance, a first-order transition comes with a characteristic two-step relaxation to equilibrium which emerges as one approaches the transition from the ergodic phase. 

\section{Hyperbolic lattices} \label{sec:hyperbolic_lattices}

\subsection{General description} \label{sub:general_description}

\begin{figure}
	[htbp] 
	\begin{center}
		\includegraphics[draft=false,width=8cm]{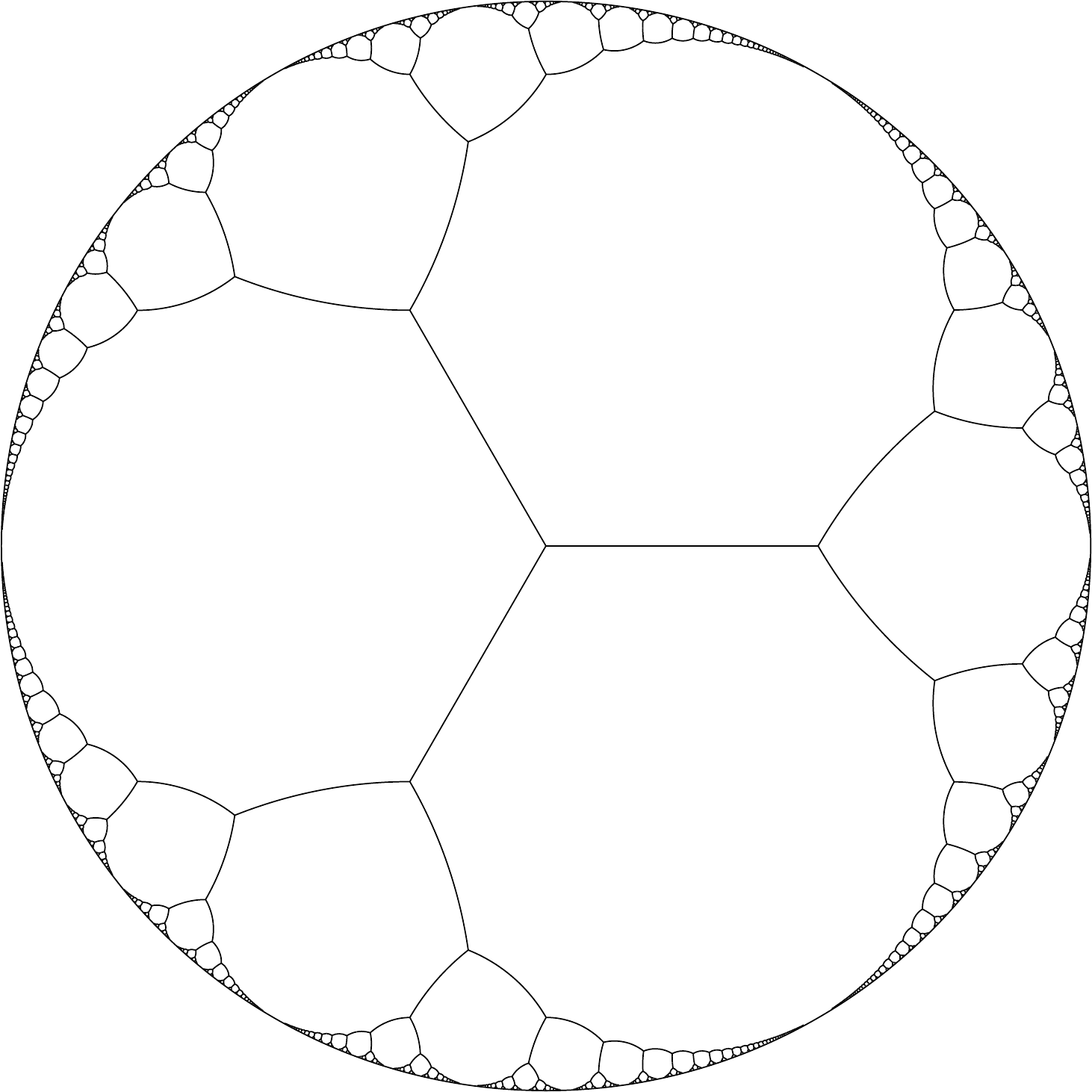} 
	\end{center}
	\caption{$\{\infty,3\}$ hyperbolic tiling in the Poincaré disk representation of the hyperbolic plane. The graph formed by the vertices and the edges of the tiling has a tree structure. Note the presence of infinite loops which are the discrete analogs of horocircles (see Appendix~\ref{sec:existence_of_finite_blocked_clusters}). If periodic boundary conditions are employed, this lattice can then be considered as a Bethe lattice.} \label{fig:Bethe} 
\end{figure}

The hyperbolic plane $H^2$, also called pseudosphere or Bolyai-Lobatchevski plane, is a Riemannian surface of constant negative curvature~\cite{Hilbert:1983,Coxeter:1969}. Contrary to a sphere, which is a surface of constant positive curvature, $H^2$ is infinite and cannot be embedded as a whole in the three-dimensional Euclidean space. ``Models'', \textit{i.e.} projections, must thus be used for its visualization. The hyperbolic metric in polar coordinates $(r, \phi)$ is given by 
\begin{equation}
	\label{eq:metric} \mathrm{d}s^2= \mathrm{d}r^2 + \left(\frac{\sinh(\kappa r)}{\kappa} \right)^2 \mathrm{d}\phi^2 
\end{equation}
with $-\kappa^2<0$ the Gaussian curvature. In the following, to represent hyperbolic lattices, we use the Poincaré disk model that maps the whole infinite space $H^2$ onto the open disk of radius unity. This projection ($r'=\tanh(\kappa r/2), \phi'=\phi$) is conformal, \textit{i.e.} it preserves the angles, but is not isometric: the Euclidean distance between two points of the disk separated by a given distance in $H^2$ shrinks to zero when the points approach the disk perimeter (see Figure~\ref{fig:Bethe}).

In this work, we consider hyperbolic lattices resulting from regular tilings or ``tessellations'' of the hyperbolic plane. Such tilings are generated by elements of discrete subgroups of the isometry group of $H^2$, also called Fuchsian groups~\cite{Poincare:1882}. Each regular tessellation is characterized by the number $p$ of faces (\textit{i.e.} edges in two dimensions) of the primitive cell and the number $q$ of adjacent cells meeting at any vertex of the tiling; they will be denoted by the Schläfli symbol $\{p,q\}$. In the hyperbolic plane, regular tessellations exist provided 
\begin{equation}
	\label{eq:tiling} (p - 2) (q - 2) > 4, 
\end{equation}
which allows an infinite number of possibilities~\cite{Coxeter:1969,Hilbert:1983,Coxeter:1965}. The integers $p$ and $q$ can thus vary between $3$ and $\infty$. Note that the cases in which either $p$ or $q$ is infinite (but not both) are well defined: for instance, the lattice $\{\infty,q\}$ is a tree of connectivity $q$ (see Figure~\ref{fig:Bethe}). 

The negative curvature of the space in which these lattices are embedded imposes strong geometrical constraints that give rise to quite different properties than those encountered in the more familiar Euclidean lattices. First, once $\{p,q\}$ have been fixed, the length of the polygon sides is also fixed and it is at least of the order $\kappa^{-1}$~\cite{Beardon:1983}: this is why kinetically constrained models on hyperbolic lattices can only explore length scales above $\kappa^{-1}$. Furthermore, the metric of the hyperbolic plane given in Equation~\eqref{eq:metric} induces that for a hyperbolic disk of radius $r$ in $H^2$, the ratio of its perimeter to its area is given by $\frac{2 \pi \kappa^{-1}\,\sinh(\kappa r)}{4 \pi \kappa^{-2} \,\sinh^2\left( \frac{\kappa r}{2} \right)}$ and goes to a nonzero value, $\kappa$, when $\kappa r \rightarrow \infty$. Thus, surface (boundary) effects can never be neglected in the hyperbolic plane. The same reasoning carries over to hyperbolic lattices which therefore belong to the class of \emph{nonamenable} graphs~\cite{Lyons:2000}: graphs with a nonzero ``edge-isoperimetric constant'', \textit{ i.e.} such that the number of edges between vertices of the graph and the exterior scales like the number of interior edges. Cayley trees are examples of nonamenable graphs (but random graphs are amenable). Nonamenable graphs share common properties~\cite{Lyons:2000}: they are in some sense infinite-dimensional and lead in many cases to phase transitions having mean-field character~\cite{Schonmann:2002}.

\subsection{Relation with trees} \label{sub:relationship_with_trees}

In this section we describe tree-like structures which can be embedded into hyperbolic lattices and will be useful in the analysis of the BP transition. For all hyperbolic lattices with $p>4$, by cutting appropriate links one can construct spanning trees with vertices of connectivity $q$ and $q-1$. The general procedure to build such spanning trees will be detailed in ~\cite{toninelli09}. It relies on  a modification of the procedure devised in~\cite{Margenstern:2006} to build trees that cover the hyperbolic lattices but contain additional edges which are not present in the original lattice\footnote{Our modification guarantees that vertices which are neighbors on the tree are necessarily neighbors on the lattice, unlike the tree resulting from the procedure in~\cite{Margenstern:2006}. This is done as follows: whenever there are $j$ links that connect a polygon of generation $i$ to polygons of generation $i+1,\dots,i+j$ in the spanning trees of~\cite{Margenstern:2006}, we substitute the $l$-th link with a link between the polygon of generation $i+(l-1)$ and the one of generation $i+l$.}. Finally, note that for $p=4$ a similar procedure exists but the connectivity of vertices is $q-1$ and $q-2$~\cite{Margenstern:2003}.

\begin{figure}
	[htbp] 
	\begin{center}
		\subfloat[]{\label{fig:6-5_loops} 
		\includegraphics[draft=false,width=6cm, trim = 7cm 0 0 7cm, clip = true]{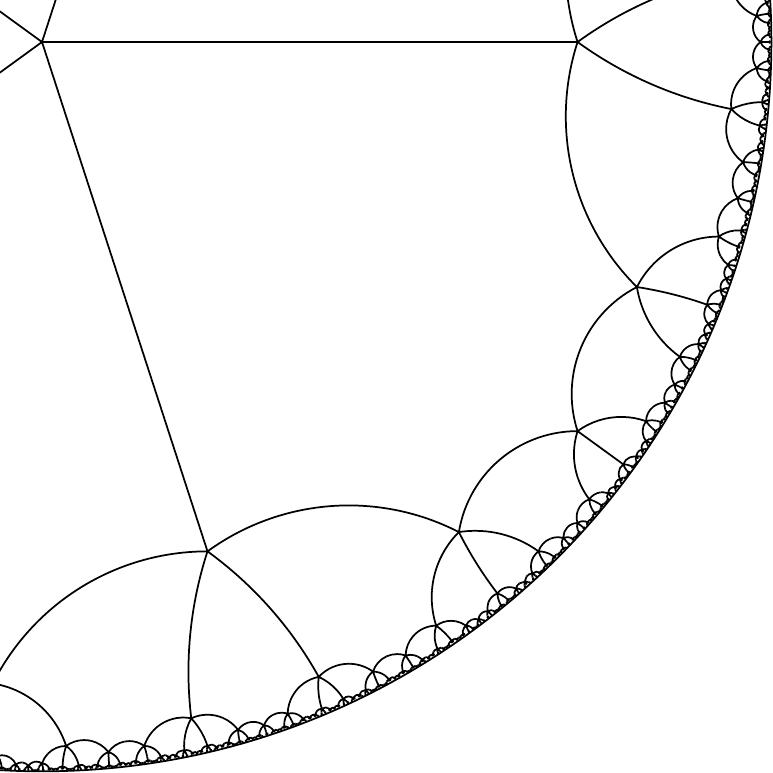}} \hspace{0.5cm} \subfloat[]{\label{fig:6-5_tree} 
		\includegraphics[draft=false,width=6cm, trim = 7cm 0 0 7cm, clip = true]{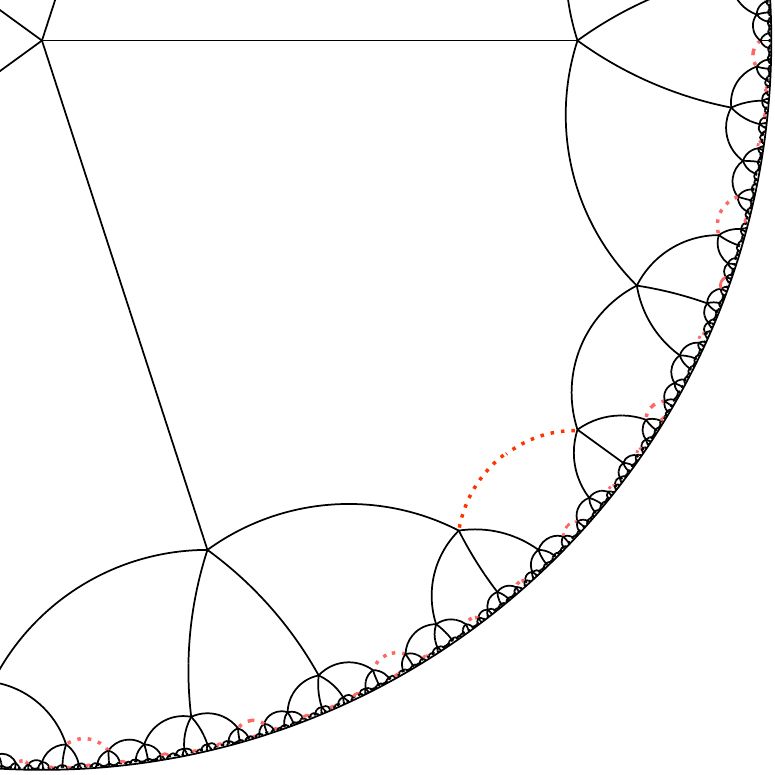}} 
	\end{center}
	\caption{ \label{fig:6-5} Part of the $\{6,5\}$ hyperbolic tiling in the Poincaré disk representation of the hyperbolic plane: \ref{fig:6-5_loops} full lattice with loops; \ref{fig:6-5_tree} spanning tree with discarded edges shown as red dotted lines. Note the small number of edges that one needs to cut to avoid loops.} 
\end{figure}

The existence of these spanning trees with mixed connectivity also implies that a Cayley tree of constant connectivity $q-1$ (respectively, $q-2$) obtained by cutting some additional edges in the above spanning trees can be embedded in a hyperbolic lattice $\{p,q\}$ for $p>4$ (respectively, $p=4$). This regular tree is not spanning but one can easily build a collection of such regular trees which are overlapping and whose union covers the whole lattice. An example of spanning tree for the $\{6,5\}$ lattice (which will be studied in detail in the following) is shown in Figure~\ref{fig:6-5_tree}. Note that vertices are not all equivalent: there are two types of vertices with connectivity $5$ (depending on whether one or two of their direct descendants have connectivity $4$) and one type with connectivity $4$, as shown in Figure~\ref{fig:spanning-tree}. 

\section{Existence of a bootstrap percolation transition on hyperbolic lattices} \label{sec:bootstrap_existence}

\subsection{Sketch of the proof} \label{sub:sketch_of_proof}

Compared to standard bond and site percolation that has been quite thoroughly studied on hyperbolic lattices and nonamenable graphs~\cite{Lalley:1998,Lalley:2001,Benjamini:2001,Schonmann:2002,Baek:2009}, BP has received less attention. It has been solved on Bethe lattices and Cayley trees~\cite{Chalupa:1979} and the result $\rho_c<1$ has also been proved on other nonamenable graphs under appropriate conditions on the graph topology~\cite{Balogh:2006}. Here, we adopt a strategy that is specifically devised for hyperbolic lattices and we consider generic values of $m$ and of the lattice parameters $p$ and $q$. Let us start by stressing that even if the hyperbolic lattices $\{p,q\}$ are nonamenable, they have loops (cycles) on all length scales for finite $p$ and $q$. Thus they share at the same time properties of trees (nonamenability) and of Euclidean lattices (loops on all length scales). It is therefore \textit{a priori} unclear if a BP transition at a nontrivial density should be expected as on trees or if the presence of finite loops wipes out the BP transition as in Euclidean lattices.

For some choices of the parameters $m,p$ and $q$ ($m=2$ whatever $p,q$ and $m=3, p=3$ whatever $q$, see Appendix~\ref{sec:existence_of_finite_blocked_clusters}), there exist finite blocked clusters for any $\rho>0$. These cases are not very interesting from the physical point of view since the corresponding FA model is nonergodic at any nonzero density (\textit{i.e.}, trivially, $\rho_c=0$), whereas an ergodicity-breaking transition only makes physical sense if it is due to an infinite cluster of jammed particles. We disregard these cases and focus on lattices characterized by $p>3, m>2$ and $p=3,m>3$. As explained in Appendix~\ref{sec:existence_of_finite_blocked_clusters}, it is only in these cases that blocked clusters are necessarily infinite. In addition, we do not consider the case $m=q$ ($f=1$) which leads to an all empty configuration as soon as there exists at least one empty site in the initial configuration (thus, trivially, $\rho_c =1$).

We now state some basic results that will be useful for the analysis of BP with blocking parameter $m$ on a graph $G$. Consider a subgraph $\tilde{G}\subset G$, then the following inequality holds between the BP thresholds: 
\begin{equation}
	\label{eq_upper_bound} \rho_c(G,m) \leq \rho_c(\tilde{G},m). 
\end{equation}
The reason is that a blocked configuration on $\tilde{G}$ is also blocked on $G$ because adding links cannot help unblocking a structure: the number of blocked neighbors in $G$ can only increase or remain equal to the value in $\tilde{G}$. The other important inequality that will be needed in the following concerns $\overline\rho_c(G,f)$, the critical density on the graph $G$ with the facilitation parameter $f$.

If instead $\tilde G$ contains all the vertices and a subset of the edges of $G$ 
and we perform BP with constant facilitation parameter $f$ we obtain 
\begin{equation}
	\label{eq_lower_bound} \overline\rho_c(G,f) \geq \overline\rho_c(\tilde{G},f). 
\end{equation}
The reason is that an unblocked configuration on $\tilde{G}$ is also unblocked on $G$ because adding links can only increase or leave constant the number of empty neighbors of a given site. A direct consequence of this result is that if we have a collection of (possibly overlapping) graphs $\tilde G_i\subset G$ such that their union $\cup_i\tilde G_i$ covers $G$, then it analogously follows that 
\begin{equation}
	\label{eq_lower_bound3} \overline\rho_c(G,f) \geq \inf_i \, \overline\rho_c(\tilde{G}_i,f). 
\end{equation}

Finally, we note that on lattices with a fixed connectivity $q$, as for the hyperbolic lattices, $\overline\rho_c(G,f)=\rho_c({G},m(f))=\rho_c({G},q+1-f)$. As a consequence, the two inequalities \eqref{eq_lower_bound3} and \eqref{eq_upper_bound} allow us to obtain lower and upper bounds on the critical BP density on hyperbolic lattices.

As explained in section~\ref{sub:relationship_with_trees}, there exists for the hyperbolic lattice $\{p,q\}$ with $p>4$ a sequence of Cayley trees $\tilde G$ of constant connectivity $\tilde{q}=q-1$ that covers the lattice. Thus, we can apply \eqref{eq_upper_bound} and \eqref{eq_lower_bound3} to BP with blocking parameter $m$ on the lattice $\{p,q\}$, noticing that $\overline \rho_c(\tilde{G},f)=\rho_c(\tilde{G},\tilde{q}-f+1)=\rho_c(\tilde{G},m-1)$ (recall that the facilitation parameter for a hyperbolic lattice $\{p,q\}$ is $q+1-m$). Hence, we obtain: 
\begin{equation}
	\label{eq:encadrement} \rho_c(T^{q-1},m-1) \leq \rho_c(\{p,q\},m) \leq \rho_c(T^{q-1},m), 
\end{equation}
where we have used the notation $T^q$ to indicate an infinite tree of constant connectivity $q$. As we have already recalled in Section~\ref{sub:KCM_and_bootstrap}, if $2 \leq m < q-1$, then $0 < \rho_c(T^{q-1},m) < 1$ (see \textit{e.g.}~\cite{Chalupa:1979}). Therefore, we conclude that a BP transition also occurs on $\{p,q\}$ hyperbolic lattices with $p>4$, $q>4$ for $3 \leq m < q-1$ since we obtain $0 < \rho_c(T^{q-1},m-1) \leq \rho_c(\{p,q\},m) \leq \rho_c(T^{q-1},m) < 1$. The case $m=q-1$ requires special care in order to prove that the critical density is bounded away from one. We will give a specific example (m=4, q=5, p=6) of how to prove this result in the next section. Analogously, by using the embedded trees of connectivity $q-2$ for the case $p=4$ one immediately obtains that $0<\rho_c<1$ when $4\leq m\leq q-2$. The other cases will be treated in a forthcoming publication~\cite{toninelli09}.

By similar arguments and using the spanning trees with mixed connectivities $q$ and $q-1$, one can establish tighter bounds on the critical density as will be illustrated in the next section for the $\{6,5\}$ lattice. 
\begin{figure}
	[htbp] 
	\begin{center}
		\includegraphics[draft=false,height=8cm]{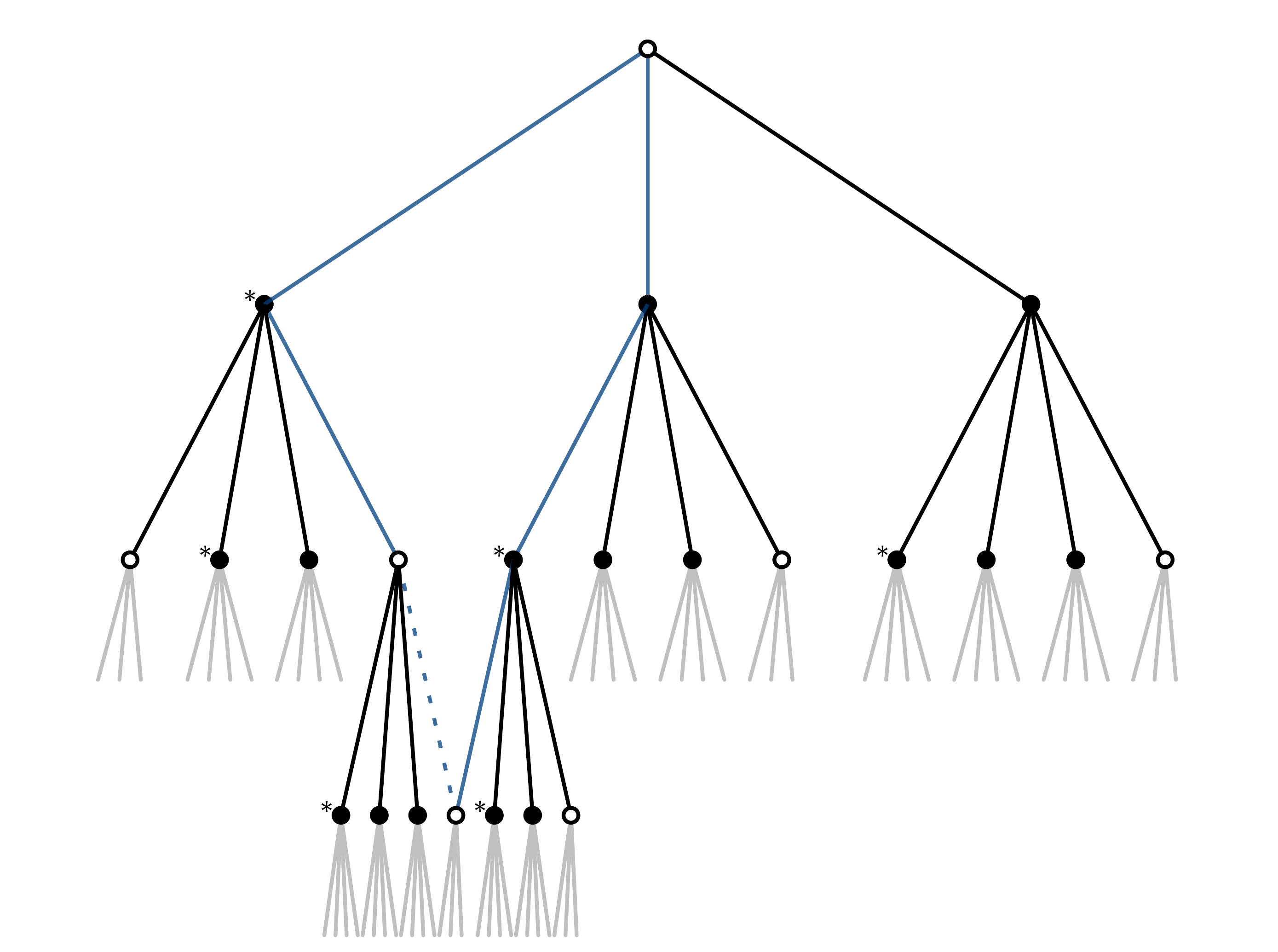} 
	\end{center}
	\caption{\label{fig:spanning-tree} Part of a tree with mixed connectivity $(5,4)$ that spans the $\{6,5\}$ hyperbolic lattice. Edges indicated by the dotted lines correspond to branches of the original lattice pruned to avoid loops. The blue edges correspond to an hexagon of the original $\{6,5\}$ lattice. In this spanning tree, vertices are not equivalent and three different types are present, which depend on their descendants: $\circ$ possesses three descendants ($^*\bullet,\bullet,\bullet$), $\bullet$ has four descendants ($^*\bullet,\bullet,\bullet,\circ$) and $^*\bullet$ has four descendants ($\circ,{^*\bullet},\bullet,\circ$). Note that the top vertex has also one parent, not shown in the figure.} 
\end{figure}

\subsection{Illustration: the $\{6,5\}$ lattice with $m=3$} \label{sub:illustration}

For the $\{6,5\}$ lattice, a possible spanning tree with mixed connectivity $4$ and $5$, which will be denoted by $T^{4,5}$, is shown in Figures~\ref{fig:6-5_tree} and~\ref{fig:spanning-tree}. The construction rules follow the procedure of~\cite{Margenstern:2006} with the modification explained in section~\ref{sub:relationship_with_trees} (see caption of Fig.~\ref{fig:spanning-tree}).

We now derive $\rho_c(T^{4,5},3)$ and $\overline\rho_c(T^{4,5},3)$ which, thanks to inequalities~\eqref{eq_upper_bound} and~\eqref{eq_lower_bound}, respectively provide an upper and a lower bound to the critical density $\rho_c(\{6,5\},3)=\overline\rho_c(\{6,5\},3)$ of the hyperbolic lattice under consideration.

As explained in the caption of the Figure~\ref{fig:spanning-tree}, the vertices of the tree are of three different types, corresponding to two different connectivities. Thanks to the tree structure and following the same method as in~\cite{Sellitto:2005,Toninelli:2005}, one can write for each type of vertex a recurrence relation on the probability $P$ that a site is blocked assuming that its ``parent'' vertex is itself blocked. For instance, by considering a vertex of type $\circ$ (see Figure~\ref{fig:spanning-tree}), one can derive its probability $P_{\circ}$ as a function of the probabilities of its descendants which are of types $^*\bullet$ and $\bullet$, \textit{i.e.} $P_{^*\bullet}$ and $P_{\bullet}$. The procedure can be repeated for vertices of the two other types and this provides equations relating the probabilities $P_{\circ},P_{^*\bullet}$ and $P_{\bullet}$ at two successive generations of the tree. Then, we perform the standard ``Bethe lattice'' trick which consists of considering only the deep interior of the tree and assuming that all vertices of the same type are equivalent \footnote{When we refer to the Bethe lattice, we think of an infinite tree-like structure as is the case for an infinite Cayley tree. However, while the latter is constructed via the infinite volume limit of a finite tree and therefore has a boundary at infinity (i.e. leaves), the Bethe lattice has no boundary.}. As can be shown on the Cayley tree, this procedure that amounts to avoiding boundary effects does not affect the location of the threshold at $\rho_c$ (see below for a more detailed discussion). This leads to the following set of coupled equations for BP with blocking parameter $m=3$: 
\begin{equation}
	\begin{cases}
		P_{\circ} = \rho \; P_{\bullet} \left( P_{\bullet} + 2 P_{^*\bullet}\left( 1- P_{\bullet} \right) \right) \\
		P_{^*\bullet} = \rho \left( P_{\circ}^2 + 2 P_{\circ} (1-P_{\circ}) (P_{^*\bullet}+P_{\bullet}) + P_{^*\bullet}P_{\bullet}(1-4 P_{\circ} + 3 P_{\circ}^2) \right) \\
		P_{\bullet} = \rho \left( P_{\bullet}^2 + 2 P_{\bullet} (1-P_{\bullet}) (P_{^*\bullet}+P_{\circ}) + P_{^*\bullet}P_{\circ}(1-4 P_{\bullet} + 3 P_{\bullet}^2) \right), 
	\end{cases}
\end{equation}
and for BP with facilitation parameter $f=3$: 
\begin{equation}
	\begin{cases}
		P_{\circ} = \rho \left( 1 - (1-P_{^*\bullet})(1 - P_{\bullet})^2 \right) \\
		P_{^*\bullet} = \rho \left( P_{\circ}^2 + 2 P_{\circ} (1-P_{\circ}) (P_{^*\bullet}+P_{\bullet}) + P_{^*\bullet}P_{\bullet}(1-4 P_{\circ} + 3 P_{\circ}^2) \right) \\
		P_{\bullet} = \rho \left( P_{\bullet}^2 + 2 P_{\bullet} (1-P_{\bullet}) (P_{^*\bullet}+P_{\circ}) + P_{^*\bullet}P_{\circ}(1-4 P_{\bullet} + 3 P_{\bullet}^2) \right). 
	\end{cases}
\end{equation}
A numerical solution of the above two systems of equations yields in both cases the coexistence of two solutions above a critical density smaller than one, thus for our hyperbolic lattice via inequalities \eqref{eq_upper_bound} and~\eqref{eq_lower_bound} we get
\begin{equation}
	\rho_c(T^{4,5},3) \simeq 0.711 \leq\rho_c(\{6,5\},3)\leq \overline\rho_c(T^{4,5},3)\simeq 0.734, 
\end{equation}
which provides a rather good estimate of the critical density for the hyperbolic lattice under consideration. (The simpler bounds of Equation~\eqref{eq:encadrement} on the other hand provide $\rho_c(T^{4},2)=1/3 \leq\rho_c(\{6,5\},3)\leq \overline\rho_c(T^{4},3)=8/9$.)

For $m=4$, the same method can be applied and the numerical solution of the corresponding equations yields $0.940\leq\rho_c(\{6,5\},4)\leq 0.985$. Thus, a transition also takes place for $m=4$ in the $\{6,5\}$ lattice. Note that this gives one specific example of how to deal with the case $m=q-1$ for which the existence of a transition at a finite critical density cannot be proved by simply using the spanning collection of regular trees of connectivity $q-1$.

\section{Order of the transition} \label{sec:bootstrap_order}

In the previous section, we have established the existence of a BP transition on appropriate hyperbolic lattices at $0<\rho_c<1$. Here we provide evidence supporting that this transition has a first-order (\textit{i.e.} discontinuous) character: the order parameter which corresponds to the probability that a given site is blocked is discontinuous at $\rho_c$. We also obtain numerical indication of a diverging correlation length. Thus, the behavior resembles that of the BP transition on tree-like structures such as Bethe lattices and random graphs.

Unfortunately, the analysis of the first-order character of the BP transition is much more involved for hyperbolic lattices than for either Bethe or Euclidean lattices. In the latter cases, one can simply measure the distribution of blocked sites. For a discontinuous transition, this would show a characteristic double-peaked structure, with the peak centered on zero corresponding to the absence of any blocked cluster and the other peak centered at a finite distance from zero corresponding to the appearance of a finite fraction of blocked sites at the transition. This strategy does not work for hyperbolic lattices because of the effect of boundary sites. Due to the nonamenability of the lattices, one has to be very careful with boundary effects, as we explain in the following.

\subsection{Taking into account boundary condition effects: discussion and some conjectures} \label{sub:conjectures}

\begin{figure}
	[htbp] 
	\begin{center}
		\includegraphics[draft=false,height=7cm]{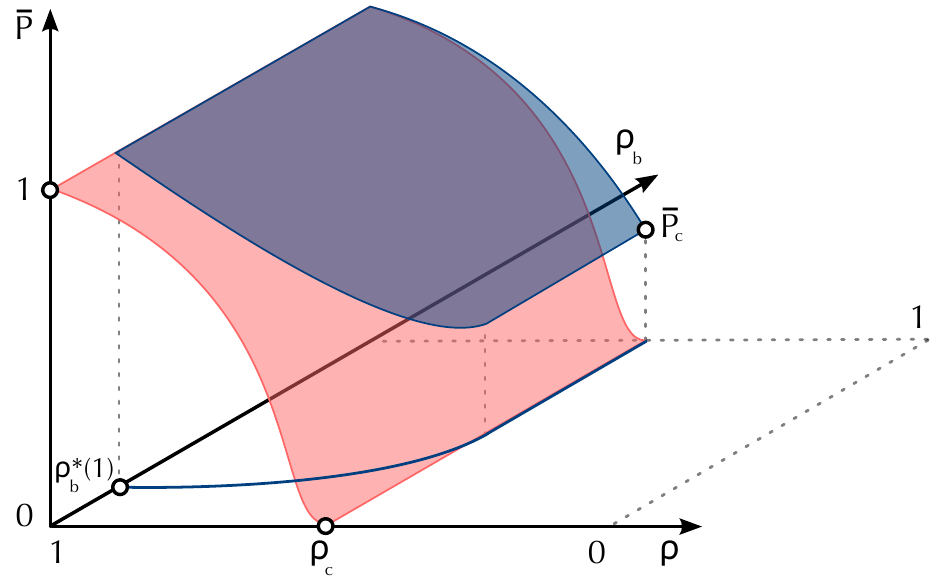} 
	\end{center}
	\caption{Probability $\overline{P}$ for a vertex to be blocked depending on the initial density $\rho$ and the density at the boundary $\rho_{b}$ for a continuous (in red) and a discontinuous (in blue) BP transition. $\rho_c$ is the density at the transition, $\overline{P_c}$ the nonzero probability for a vertex to be blocked at the transition when the latter is discontinuous, $\rho_{b}^*(\rho)$ (the blue line in the $\overline{P}=0$ plane) the boundary density below which the system is completely emptied for all values of the initial bulk density $\rho$, and $\rho_{b}^*(1)=\rho_{b}^*(\rho=1)$. The discontinuous data are for a Cayley tree of connectivity $q=5$ with $m=3$: then, $\rho_c = \frac{68 - 5 \sqrt{10}}{72} \simeq 0.7248 $ and $\overline{P_c} \simeq 0.4132 $. Note that the direction of the $\rho$ axis is reversed to improve the readability of the figure.} \label{fig:order} 
\end{figure}

We consider a hyperbolic lattice and a disk of large radius $R$ centered on a given vertex of the lattice which we refer to as the origin. We call \textit{boundary sites} those that are outside the disk but are connected to at least one vertex inside the disk. The vertices inside the disk evolve according to the BP dynamics, starting from an initial configuration with a density $\rho$ and with the boundary vertices frozen to a configuration in which sites are independently occupied with probability $\rho_{b}$. The influence of $\rho_{b}$ on the final state inside the disk after the bootstrap dynamics is a question that has not yet been addressed, even on trees. We make two conjectures on this influence for nonamenable systems which display in infinite volume a BP transition at $0< \rho_c <1$. As will be shown in the following section, the validity of these conjectures can be checked on the exactly solvable case of the Cayley tree (which is a nonamenable graph and a special case of a hyperbolic lattice). Call $\overline{P_0}(\rho,\rho_{b})$ the limit as $R\to\infty$ of the probability that the vertex at the origin is blocked and $\rho_{b}^*(\rho)$ the minimal boundary density which gives a non zero $\overline P_0$ (\textit{i.e.} $\overline P_0(\rho,\rho_{b})=0$ only if $\rho_{b}<\rho_{b}^*(\rho)$). We conjecture that 
\begin{itemize}
	\item if the BP transition is discontinuous and $\rho$ is infinitesimally above $\rho_c$ then $\rho_{b}^* > 0$ (\textit{i.e.} $\lim_{\epsilon\to 0^+}\rho_{b}^*(\rho_c+\epsilon)>0$). 
	\item if the BP transition is continuous and $\rho$ is infinitesimally above $\rho_c$, then $\rho_{b}^* =0$ (\textit{i.e.} $\lim_{\epsilon\to 0^+}\rho_{b}^*(\rho_c+\epsilon)=0$). 
\end{itemize}
Figure~\ref{fig:order} summarizes how to distinguish between continuous and discontinuous BP transitions by using the above conjectures. Note that $\rho_{b}^*$ is a monotonously decreasing function of $\rho$ since a larger initial bulk density requires the same or a lower boundary density to support the bulk blocked particles. (In Figure~\ref{fig:order} we show a discontinuous case with $\lim_{\rho\to 1}\rho_{b}^*>0$, as it happens on Cayley trees, see below. However in other cases this limit might be zero.) 

Another important remark is that for BP with periodic boundary conditions one can take as an order parameter either the fraction of blocked sites or the probability that the origin is blocked. Here, by contrast, boundary conditions are fixed and the former definition is not meaningful since the boundary may support a very large number of blocked particles in its vicinity even though the clusters of blocked particles do not percolate. In general, when dealing with nonamenable systems one should be careful because boundary effects can propagate into the entire system. Several recent studies seem to overlook this point~\cite{Shima:2006,Shima:2006a,Baek:2007,Baek:2008,Baek:2009}. To properly control these effects, one should either study them explicitly as we do here (see also~\cite{Wu:2000,Angles-dAuriac:2001,Doyon:2004,Baek:2009a}) or use periodic boundary conditions~\cite{Balazs:1986,Sausset:2007,Modes:2007,Sausset:2008b}. 

\subsection{Cayley tree} \label{sub:cayley_tree}
\begin{figure}
	[htbp] 
	\begin{center}
		\captionsetup[subfloat]{justification=raggedright,singlelinecheck=false} \subfloat[Discontinuous transition ($m>2$): $\rho_{b}^*(1)$ represents the threshold of $\rho_{b}$ below which the system is totally emptied when the density $\rho=1$; it corresponds to an unstable fixed point. $P=0$ is always a stable fixed point and an additional stable fixed point appears for $\rho > \rho_c$ at high values of $P$.]{\label{fig:discontinuous} 
		\includegraphics[draft=false,width=8cm]{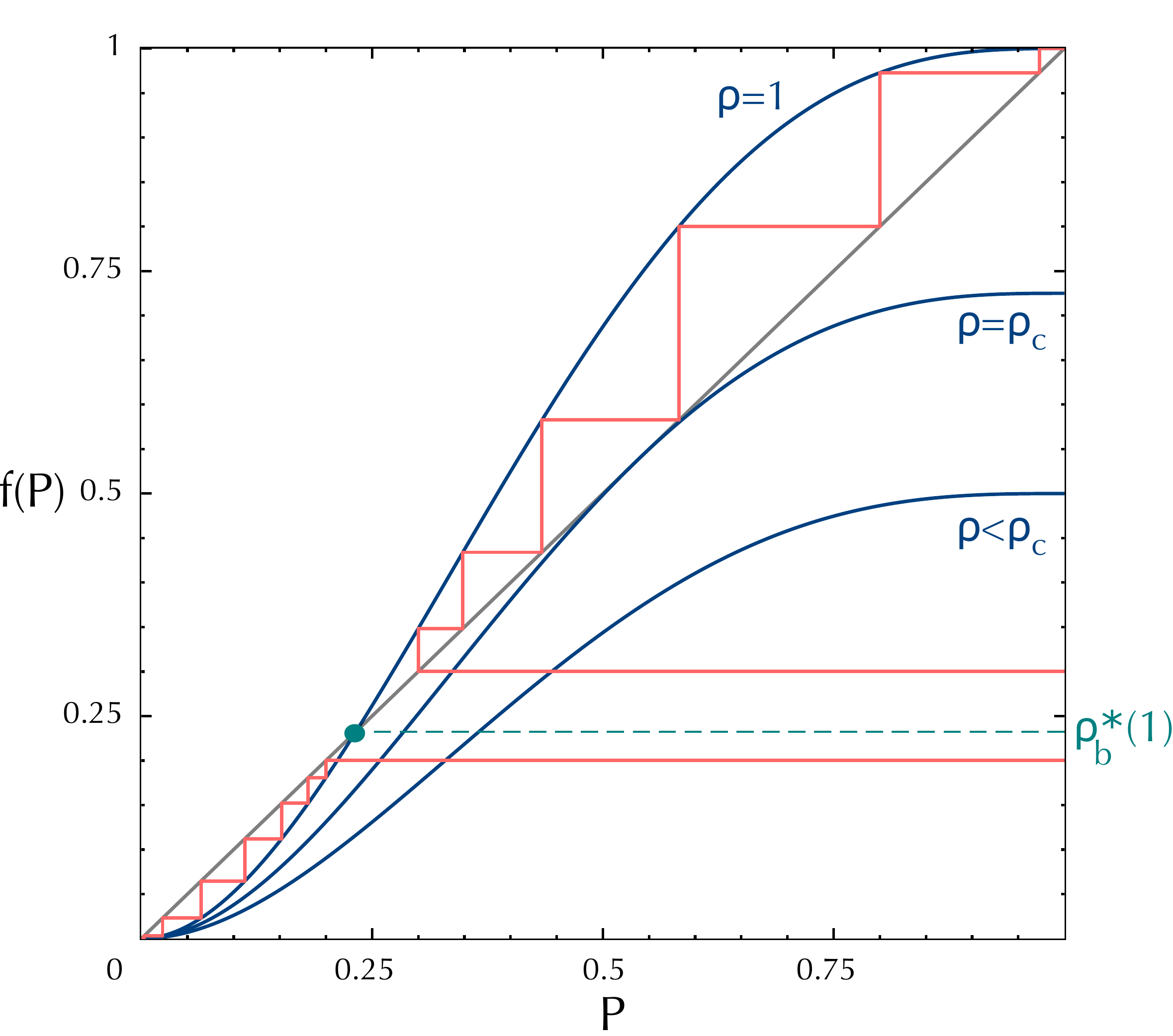}} \hspace{0.5cm} \subfloat[Continuous transition ($m=2$): there is one fixed point at $P=0$ which is stable for $\rho \leq \rho_c$ and becomes unstable for $\rho > \rho_c$. Another stable fixed point appears for $\rho > \rho_c$.]{\label{fig:continuous} 
		\includegraphics[draft=false,width=8cm]{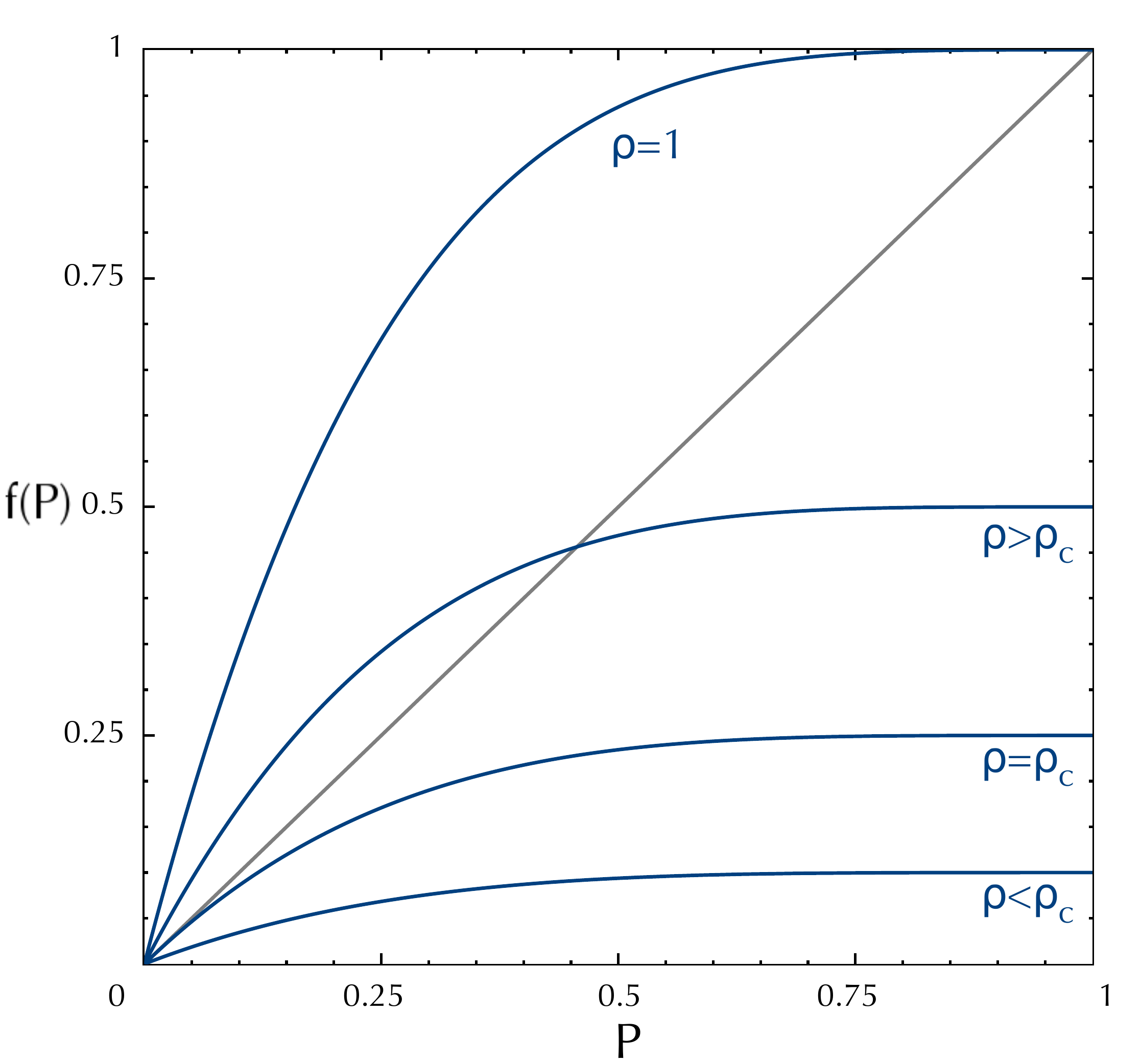}} 
	\end{center}
	\caption{Generic graphical representation of the recurrence relation for the evolution of the probability $P$ that a vertex in a Cayley tree is occupied and blocked assuming its parent is occupied and blocked; $f(P)$ is the right-hand side of Equation~\eqref{eq:recurrence}} \label{fig:tree-recurrence} 
\end{figure}

We now verify the conjectures of the previous section for a specific nonamenable graph, the Cayley tree of connectivity $q$ when $0<\rho_c<1$, \textit{i.e.} for $2\leq m\leq q-1$. Thanks to the tree structure, the following recurrence relation holds for the probability $P_i$ that a vertex of generation $i$ is blocked assuming that its parent vertex of generation $i-1$ is blocked:
\begin{equation}
	\label{eq:recurrence} P_{i} = \rho \sum_{k = 1}^{q+1-m} 
	\begin{pmatrix}
		q - 1 \\
		k - 1 
	\end{pmatrix}
	P_{i+1}^{q-k} (1-P_{i+1})^{k-1}. 
\end{equation}
If the Cayley tree is infinite, translation invariance guarantees that $P_{i+1}=P_{i}=P$ and we obtain from Equation~\eqref{eq:recurrence} a fixed point equation for $P$ (the same property applies in the Bethe lattice procedure). Furthermore, the probability $\overline{P}$ for a vertex to be blocked (without assuming its parent to be blocked) satisfies 
\begin{equation}
	\label{eq:central_site} \overline{P} = \rho \sum_{k = 0}^{q-m} 
	\begin{pmatrix}
		q \\
		k 
	\end{pmatrix}
	P^{q-k} (1-P)^{k} . 
\end{equation}
If $P$ is nonzero, then $\overline{P}$ is nonzero too and from Equation~\eqref{eq:recurrence} one can compute the critical density $\rho_c$ at which the BP transition takes place~\cite{Balogh:2006}. The transition is continuous ($P(\rho_c)=\overline P(\rho_c)=0$) for $m=2$ and discontinuous for $3<m<q-1$. 

If instead we perform the finite volume procedure with frozen boundary conditions of density $\rho_b$ (say on generation $n$) detailed above, vertices are no longer equivalent and Equation~\eqref{eq:recurrence} can be directly used to compute how $P_i$ evolves when starting from the $n$-th frozen generation. The probability $\overline{P}_0$ can be explicitly computed for any couple of values of $\rho$ and $\rho_{b}$. The outcome is plotted in Figure~\ref{fig:comparison}. We can now check that the conjectures of section~\ref{sub:conjectures} are verified.

Call $f(P)$ the function such that the right-hand side of Equation~\eqref{eq:recurrence} corresponds to $f(P_{i})$. If $3\leq m<q-1$, $f(P)$ has only one stable fixed point at $P=0$ for $\rho < \rho_c$, two fixed points at $\rho = \rho_c$, $P=0$ which is stable and $P=P_c$ which is unstable, and three fixed points for $\rho > \rho_c$, two stable ones and an unstable one, $P^u(\rho)$. In Figure~\ref{fig:discontinuous}, the evolution of $P_i$ is represented as a function of the initial condition $P_n = \rho_{b}$ for a tree with $n$ generations. If $\rho>\rho_c$ and $\rho_{b}<P^u$ then $P_0$ goes to zero, whereas if $\rho_{b}>P^u$ then $P_0$ goes to the second attractive fixed point which has a nonzero value. Thus, by using the notation of previous section, we obtain $\rho_{b}^*(\rho)=P^u(\rho)>0$ when $\rho>\rho_c$ and our conjecture on discontinuous BP transitions is verified for the Cayley tree. Note that since $P^u<P_c$ the bulk behavior on the tree is identical to the behavior in the Bethe lattice procedure considered above when $\rho_{b} \geq P_c$, while for $\rho_{b} < P_c$, boundary effects have to be taken into account (see Figure~\ref{fig:comparison}). For the opposite case $m=2$, the function $f(P)$ (see Figure~\ref{fig:continuous}) has always a fixed point at $P=0$, stable when $\rho \leq \rho_c$ and unstable otherwise, and a second stable fixed point appears for $\rho > \rho_c$. Since the stable fixed point is always unique, the bulk behavior is not affected by boundary conditions and $\rho_{b}^*=0$ for any $\rho > \rho_c$. This result for the continuous transition on the Cayley tree with $m=2$ is compatible with our conjecture in section~\ref{sub:conjectures}. 

\subsection{Generic hyperbolic lattices} \label{sub:generic_hyperbolic_lattices}

Due to the presence of loops in generic hyperbolic lattices, the iterative method used for the Cayley tree cannot be applied and we are not able to prove the conjectures of section~\ref{sub:conjectures}. However, accepting the validity of the conjectures, we obtain evidence that the transition (when it exists) is discontinuous. Indeed, we can show that there is a nonzero boundary density $\rho_b^*$ below which the system is totally emptied at the end of the BP process. 

Consider the hyperbolic lattice $\{p,q\}$ with $p>4$ and call $T^{q,q-1}$ its spanning trees of mixed connectivity $q,q-1$ (see Section~\ref{sub:relationship_with_trees}). As we did for Cayley trees in the previous section, one can verify that the BP transition on $T^{q,q-1}$ with facilitation parameter $m+q-1$ is discontinuous for $m>3$ and that $\rho_{b}^*(T^{q,q-1},\rho)>0$ for $\rho\geq\overline\rho_c(T^{q,q-1},q-m+1)$. As stressed in section~\ref{sub:sketch_of_proof}, if one performs BP with facilitation parameter $q-m+1$, each unblocked site on the spanning tree is also unblocked for BP on the $\{p,q\}$ lattice with the same facilitation parameter, \textit{i.e.} with a blocking parameter $m$. Thus, $\rho_c(\{p,q\},m)\geq\overline\rho_c(T^{q,q-1},q-m+1)$ and $\rho_{b}^*(\{p,q\},\rho)\geq\rho_{b}^*(T^{q,q-1},\rho)$. As a consequence, we get  $\rho_{b}^*(\{p,q\},\rho) \geq \rho_{b}^*(T^{q,q-1},\rho) > 0$ for $\rho>\rho_c(\{p,q\},m)$: the transition on the hyperbolic lattice $\{p,q\}$ with $p>4$ and blocking parameter $m>3$ is discontinuous provided one assumes the validity of our first conjecture in section~\ref{sub:conjectures}. In the next section we numerically analyze a case with $m=3$ which is not covered by the above argument. 

\subsection{Simulation on the $\{6,5\}$ lattice: role of boundary conditions and diverging lengthscale} \label{sub:the_{6_5}_case}

We have carried out a numerical simulation on the $\{6,5\}$ hyperbolic lattice with blocking parameter $m=3$. The objective of this study is, on the one hand, to provide additional evidence that the BP transition in hyperbolic lattices is discontinuous and, on the other hand, to unveil the existence of a diverging lengthscale when approaching the BP transition from low density.

\begin{figure}
	[htbp] 
	\begin{center}
		\includegraphics[draft=false,height=8cm]{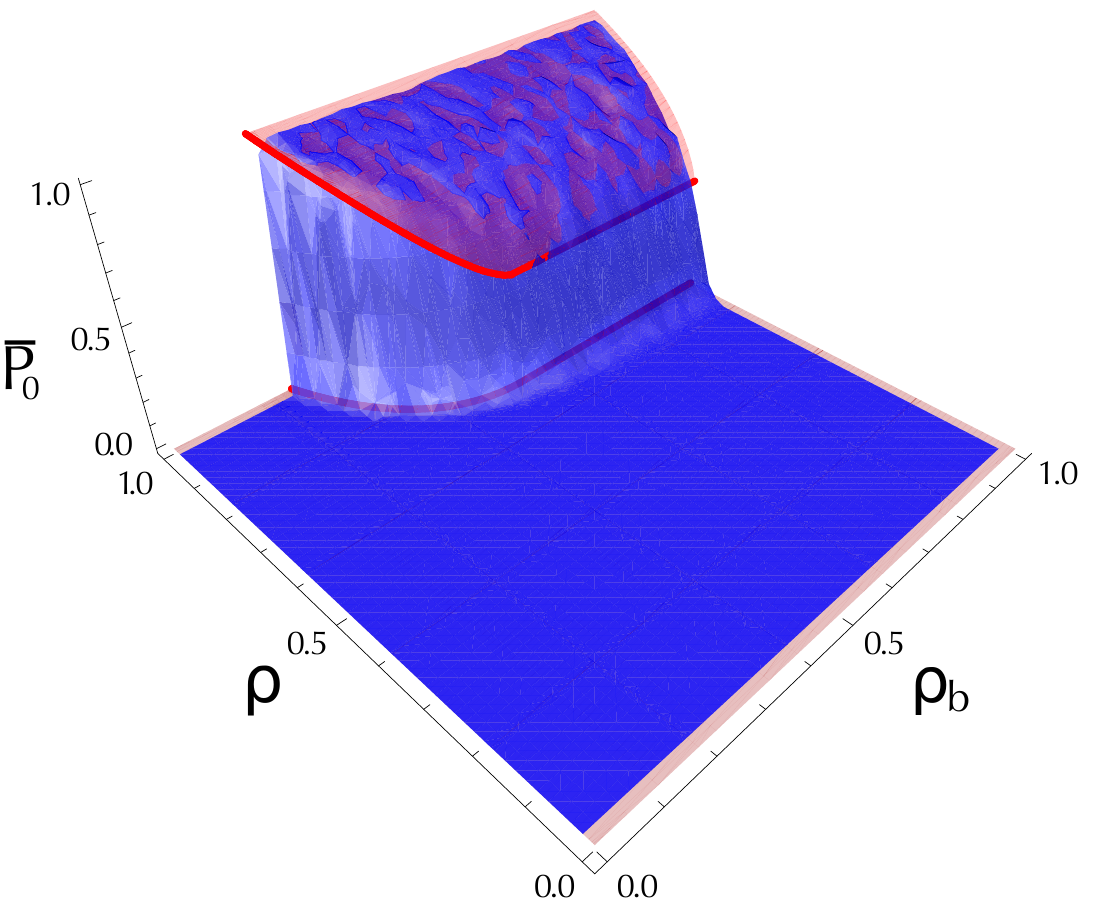} 
	\end{center}
	\caption{Probability $\overline{P}_0$ that the central vertex is occupied and blocked plotted as a function of the initial bulk and densities, $\rho$ and $\rho_{b}$. Comparison of the numerical results for the $\{6,5\}$ hyperbolic tiling (in blue) and the analytical results for the Cayley tree of connectivity $5$ (in red). The results are for a lattice of 59561 vertices and are averaged over $10^3$ different initial conditions for each value of $\{\rho,\rho_{b}\}$.} \label{fig:comparison} 
\end{figure}

Following the procedure of section~\ref{sub:conjectures}, we consider only the region inside a finite circle of given radius $R$ and the boundary sites are frozen with a density $\rho_b$. The system is initialized by filling vertices inside the circle with a density $\rho$ and the BP dynamics with $m=3$ is recursively applied until no more vertices can be emptied. Our numerical results are obtained with lattices of $27$ to $3227991$ sites, which corresponds to values of $R$ from $2$ to $14$, and are averaged over $10^3$ realizations with identical conditions $\{\rho,\rho_{b}\}$. Figure~\ref{fig:comparison} shows a three-dimensional plot of the computed probability $\overline{P}_0$ that the central site is occupied and blocked, when $\rho$ and $\rho_{b}$ vary between $0$ and $1$. 

By looking at the case $\rho = 1$, one can see that the system is completely emptied for boundary densities smaller than $\rho_{b}(1) \simeq 0.24$: thus, $\lim_{\epsilon\to 0}\rho_{b}^*(\rho_c+\epsilon)>0$ and, through the first conjecture of section~\ref{sub:conjectures}, we conclude that the transition is discontinuous.

\begin{figure}
	[htbp] 
	\begin{center}
		\includegraphics[draft=false,height=8cm]{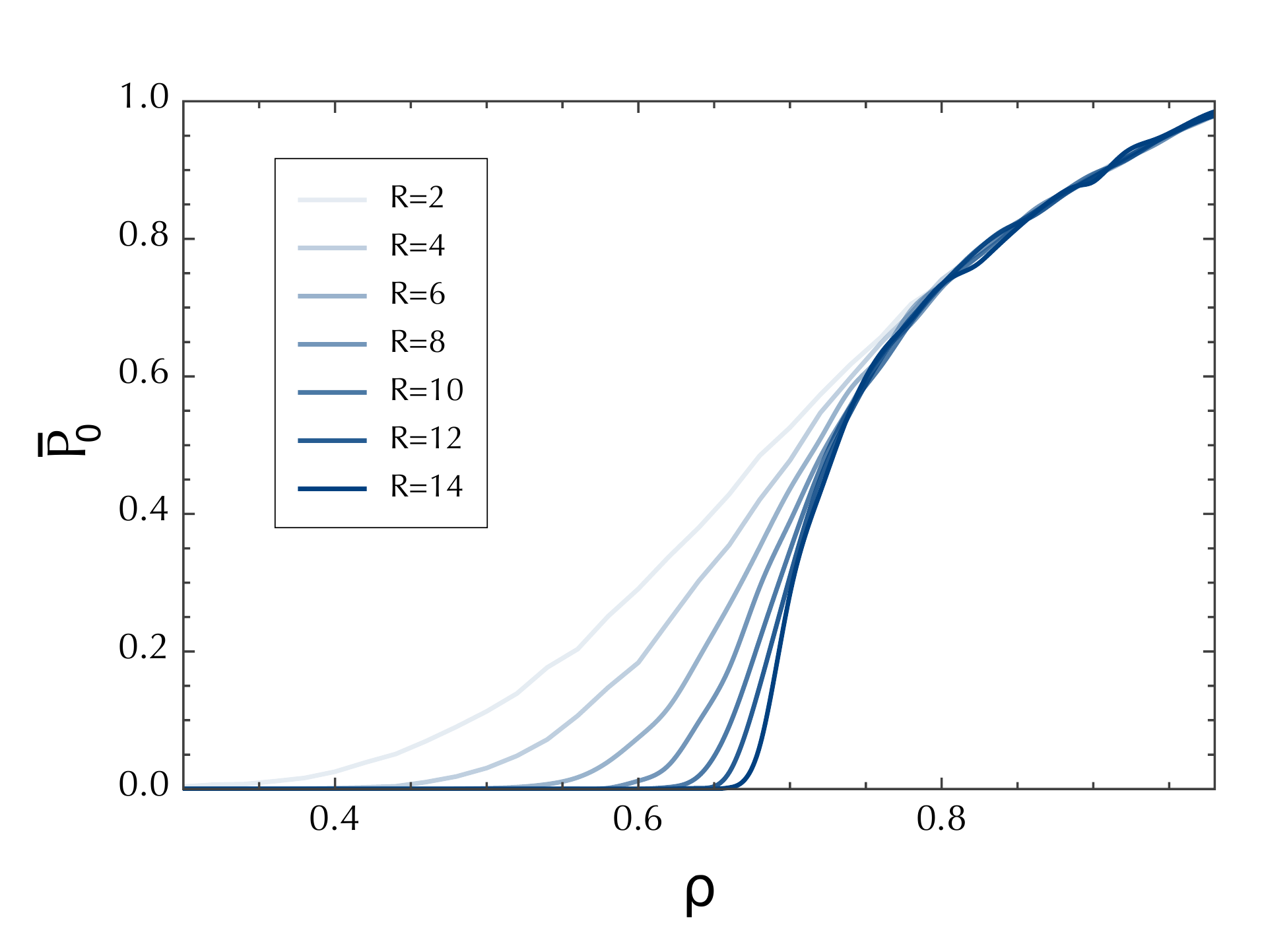} 
	\end{center}
	\caption{ Evolution of the probability $\overline{P}_0(\rho)$ for the central site to be blocked for $\rho_{b}=1$ when varying the system size $R$ between $2$ and $14$. The corresponding number of vertices goes from $27$ to $3227991$. One can note that for $\rho \gtrsim 0.75$, $\overline{P}$ is independent of $R$. } \label{fig:size-effect} 
\end{figure}

In order to support the above conclusion one should  proceed to a finite-size study of the case $\rho_{b}=1$. 
The $\rho$-dependence of the probability $\overline{P}_0$ for the central site to be blocked is displayed in Figure~\ref{fig:size-effect} for sizes ranging from $R=2$ to $R=14$. The data clearly point toward the existence of a transition at nontrivial $\rho_c \sim 0.7$, but they display strong finite-size effects which of course always make $\overline{P}_0$ a continuous function of $\rho$. 
A better description would require a finite-size scaling analysis. However the sizes we can numerically investigate are not large enough to allow a proper analysis (which is rather delicate as one knows from the corresponding study in the Cayley tree). By analogy with  the behavior on the latter graph we expect that  $\overline{P}_0(\rho)$ converges at large sizes to a function of the scaling variable $R(\rho_c-\rho)^{0.5}$. 

We stress that the finite-size effects also reveal the existence of a growing (and expectedly diverging) lengthscale when approaching the BP transition from low density, which is quite distinct from what is encountered in usual first order phase transitions. Actually, the finite-size effects indicate that for densities lower than some critical value $\rho_c$ typical initial configurations contain a finite fraction of blocked sites. It is only for system of large enough size that the boundary does not support blocked structures any more. This means that approaching the BP transition from low density in an infinite system, sub-regions of radius less than a characteristic  size  are typically blocked and can be unblocked only by starting from the boundary. Therefore, one expects that the dynamics of the corresponding FA model will show diverging dynamical correlations at the ergodicity breaking transition. In order to obtain the precise scaling of the dynamical correlations, additional work is however needed.

Finally, we note that, quite strikingly, the numerical results obtained for the $\{6,5\}$ hyperbolic lattice are very similar (up to our numerical precision and to finite size effects) to the exact ones for the Cayley tree of the \textit{same} connectivity $q=5$ (such a tree is not a subgraph of the lattice). This is illustrated in Figure~\ref{fig:comparison}. Qualitatively, the two systems appear to share the same behavior: existence of a discontinuous transition with same critical density within numerical uncertainty, same dependence with the boundary density, and same variation of $\overline{P}_0$ when $\rho > \rho_c$. The loops present (on all length scales) in the hyperbolic lattice do not seem to affect much the phenomenology of BP on hyperbolic lattices~\footnote{This conclusion certainly does not apply to the case $m=2$, for which $\rho_c=0$ on the hyperbolic lattice due to the presence of blocked structures, whereas a continuous transition takes place on Cayley trees. (To corroborate the validity of our simulations we have also verified that indeed $\rho_c \simeq 0$ for $m=2$ on the $\{6,5\}$ lattice and that in this case the bulk behavior is not influenced by the boundary density.)}. 

\section{Conclusion} \label{sec:conclusion}

In this article, we have proved that BP displays a transition at a nontrivial threshold for appropriate choices of the blocking parameter on hyperbolic lattices formed by regular tessellations of the hyperbolic plane. This in turn implies that the kinetically constrained models of the FA type have a dynamical transition at the same nontrivial threshold \footnote{By similar arguments one can also prove that a transition at a nontrivial threshold occurs on hyperbolic lattices 
for Kob Andersen models \cite{Kob:1993}, the conservative dynamics counterpart of FA models. We defer to future work a detailed analysis of these conservative dynamics}. Remarkably, this transition occurs despite the presence of loops on all length scales in hyperbolic lattices.

We have moreover provided theoretical and numerical evidence (but admittedly, not a rigorous proof) that the transition is first-order with a discontinuous jump of the order parameter. The characteristics of the transition are very similar to those on related tree structures. In particular, we have numerically shown the existence of a diverging lengthscale when approaching the BP transition from low density. This phenomenon is expected to induce diverging dynamical correlations in the dynamics of the FA model close to the ergodic-to-nonergodic transition. In fact, the dynamics of the associated FA kinetically constrained models should also bear a strong resemblance with their counterparts on Bethe lattices and random graphs, with a two-step relaxation in the ergodic phase and a jump of the nonergodicity parameter at the transition. In relation to glassforming liquids, this raises the intriguing possibility that an ideal ergodicity-breaking transition as predicted by the Mode Coupling Theory of glasses could exist for atomic liquids on the hyperbolic plane at low temperature, in a regime which is dominated by scarce topological defects.

A question that remains open for BP on hyperbolic lattices, certainly deserving further studies, is whether there exists a diverging correlation length when approaching the transition from the high-density phase. Such a correlation length has been shown~\cite{Iwata:2009,Schwarz:2006} to exist for random graphs (or Bethe lattices) and, in this context, it is related to a square root singularity of the fraction of blocked sites. Are these two phenomena also present for hyperbolic lattices? 

Finally, we conclude this work by stressing that hyperbolic lattices appear to be intermediate between random graphs and Euclidean lattices since they contain loops on all lengthscales and at the same time, because of the hyperbolic metric, are characterized by an exponential increase with the distance of the number of neighbors of a given site. It would therefore be interesting to study glassy systems and optimization problems, which have recently received much attention on random graphs~\cite{Mezard:2009}, on hyperbolic lattices. Would new phenomena take place or would the physical behavior remain the same despite the presence of loops on all lengthscales? Would the cavity techniques~\cite{Mezard:2009} developed for random graphs be useful to study glassy systems on hyperbolic lattices? We leave these questions for further investigations.
\begin{acknowledgements}
	We acknowledge partial support from the ANR Dynhet and ANRBLAN07-2184264. 
\end{acknowledgements}

\appendix

\section{Existence of finite blocked clusters} \label{sec:existence_of_finite_blocked_clusters}

First of all, it is easily realized that, whatever the values of $p$ and $q$, finite blocked clusters exist when $m=2$. Indeed, the vertices of each elementary polygon of the lattice with $p$ edges can all be blocked irrespective of the state of the vertices not belonging to this polygon. This occurs with a nonzero probability. When $p=3$, another simple case arises: for $m=3$, clusters made of $q+1$ vertices ($1$ at the center and $q$ around) are also blocked with a nonzero probability.

A simple way to find values of $m$ for which finite blocked clusters can never exist is to use a kind of ``convexity'' argument. Assume that there exists a finite blocked cluster for BP with blocking parameter $m$ on a $\{p,q\}$ lattice and consider its boundary, \textit{i.e.} vertices that are part of the cluster but connected to at least one unblocked vertex (this definition differs from that used in section~\ref{sub:conjectures}). Each boundary vertex must have at least $m$ neighbors in the cluster. The tiling being regular, the angle between two edges joining a boundary vertex to its nearest neighbors on the boundary is larger than $2\pi(m-1)/q$. If $2\pi(m-1)/q > \pi$, then the angle at each boundary vertex is strictly larger than $\pi$ and by convexity argument, the cluster must be infinite. Thus, for $m > q/2 + 1$, no finite blocked cluster can exist. One can easily check that the above arguments also hold in the case of Euclidean lattices.

We now show that the hyperbolic metric is such that finite blocked clusters exist only for $m=2$ for general $p$ and $m=3$ for $p=3$. Indeed, the bounds on $m$ can be refined. A property of the hyperbolic metric is that there exist disks of infinite radius and area, but with a boundary that does not follow a geodesic~\footnote{In the case of Euclidean geometry, a disk of infinite radius corresponds to a half space with a boundary containing geodesics. In two dimensions, the disk boundary is a line.}. Such disks are called ``horodisks'' and their boundary a ``horocircle''. One of the specificities of a horodisk is that its area is smaller than that of the half-plane whose boundary is a geodesic, contrary to the Euclidean case in which an infinite disk is equivalent to the half-plane. Obviously, clusters not contained in any horodisk are infinite.

\begin{figure}
	[htbp] 
	\begin{center}
		\includegraphics[draft=false,width=12cm]{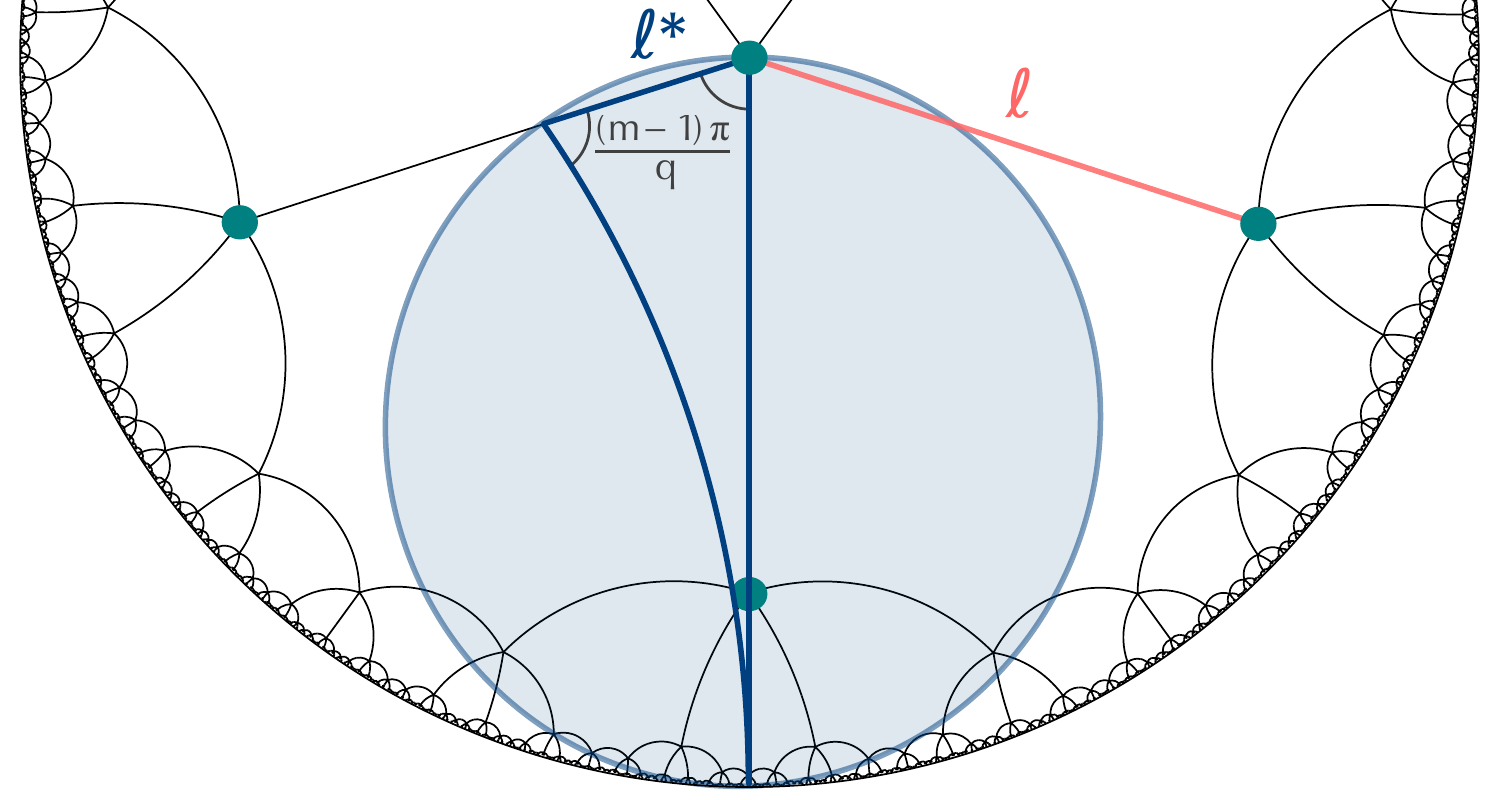} 
	\end{center}
	\caption{Geometric criterion to determine if finite blocked clusters can exist. We consider the smallest blocked cluster, which has a regular boundary with a local angle equals to $(m-1)\pi/q$ (with $m$ the number of blocking neighbors and $q$ the connectivity of the lattice). The blue disk indicates a horodisk (disk of infinite radius and infinite area) whose limiting horocircle passes through a boundary vertex of the blocked cluster (indicated by the green disk); the latter vertex possesses $m$ blocked neighbors (also indicated by green disks). $\ell^*$ corresponds to the length of the chord along the geodesic between the chosen vertex on the horocircle and one of its two neighbors on the cluster boundary. This length has to be compared with that of the lattice edges $\ell$. In the present case ($m=3$ on a $\{6,5\}$ lattice), $\ell^* < \ell$ and no finite blocked clusters exist. Note that the angle at the bottom of the blue triangle inscribed in the horocircle is zero, a specificity of hyperbolic geometry.} \label{fig:horocircle} 
\end{figure}

A geometric criterion sketched in Figure~\ref{fig:horocircle} gives a condition on $m$ for minimal clusters to be infinite, \textit{i.e.} not contained in any horodisk. For a given $m$, the smallest possible blocked cluster has a regular boundary made of blocked vertices connected to $2$ blocked neighbors on the boundary, $m-2$ blocked vertices inside the cluster and $q-m$ vertices outside the cluster. We do not consider the feasibility of such a cluster, but only whether it could be finite or not. Take any boundary vertex of the cluster. For the cluster to be finite, it must at least be contained in one horocircle passing through this boundary vertex. However, this is only possible if the chord length $\ell^*$ along the geodesic joining the chosen vertex to one of its neighbors on the cluster boundary is larger than the edge length $\ell$ in the $\{p,q\}$ lattice (see Figure~\ref{fig:horocircle}). By using the fact that the cluster boundary is regular and by means of elementary hyperbolic trigonometry (setting the curvature to $-1$), one finds 
\begin{equation}
	\label{eq_chord} \cosh(\ell^*) = \frac{\cos^2\left(\frac{(m-1)\pi}{q}\right) + 1}{\sin^2\left(\frac{(m-1)\pi}{q}\right)} 
\end{equation}
and 
\begin{equation}
	\label{eq_edge} \cosh(\ell) = \frac{\cos^2\left(\frac{\pi}{q}\right) + \cos\left( \frac{2\pi}{p} \right)}{\sin^2\left(\frac{\pi}{q}\right)}. 
\end{equation}
Summarizing the above discussion, we conclude that if $\ell^* < \ell$, or equivalently $\cosh(\ell^*) < \cosh(\ell)$, no finite blocked clusters can exist. Through Equations~\eqref{eq_chord} and~\eqref{eq_edge}, one obtains that for $p>3$, this inequality is verified for all values of $m$ such that $2 < m \leq (q-1)/2 + 1$ and for $p=3$ it is verified as soon as $3 < m \leq (q-1)/2 + 1$. No finite blocked clusters can therefore exist for these values of $m$. One can note that values of $m$ for which this inequality does not hold correspond to those for which finite blocked clusters (with nonzero probability) can easily be found (see above).

To sum up: on a $\{p,q\}$ hyperbolic lattice, finite blocked clusters exist only when $m=2$, for all values of $(p,q)$, and when $m=3$, for $p=3$ irrespective of the value of $q$. 

\bibliographystyle{spmpsci} 
\bibliography{KCM_Hyp}

\end{document}